\documentclass[lettersize,journal]{IEEEtran}
\usepackage{amsmath,amsfonts}
\usepackage{algorithmic}
\usepackage{algorithm}
\usepackage{array}
\usepackage[caption=false,font=normalsize,labelfont=sf,textfont=sf]{subfig}
\usepackage{textcomp}
\usepackage{stfloats}
\usepackage{url}
\usepackage{verbatim}
\usepackage{graphicx}
\usepackage{cite}

\begin{document}

\title{Multi-Antenna Systems by Transmissive Reconfigurable Meta-Surface}

\author{Zhendong~Li,~Wen~Chen,~\IEEEmembership{Senior~Member,~IEEE,}~Chong~He,~\IEEEmembership{Member,~IEEE,}~Xudong~Bai,~\IEEEmembership{Member,~IEEE,}~and~Jianmin~Lu,~\IEEEmembership{Member,~IEEE.}
}

%

\maketitle

\begin{abstract}
Reconfigurable meta-surface (RMS) is proposed as a very promising and novel technology, which is composed of a large number of low-cost passive elements, and can achieve passive beamforming by controlling the amplitude and phase of incident electromagnetic (EM) waves. Therefore, in order to solve the challenges of high power consumption and high cost of existing base stations (BSs), we propose a low-cost and low-power consumption transmissive RMS multi-antenna system in this paper. Specifically, we first provide an overview of the transmissive RMS multi-antenna system, including its advantages, network architecture, transmission mechanism, modulation principle, channel model and channel estimation technique. Then, we address transceiver design and optimization for downlink (DL) and uplink (UL), and some numerical results are also given to verify the effectiveness of the proposed algorithm. Finally, several potential research directions of the transmissive RMS multi-antenna system are given to inspire further investigation in future work.
\end{abstract}


\section{Introduction}
\IEEEPARstart{W}{ith} the evolution and development of mobile communication technology, the power consumption and cost of base stations (BSs) are increasing. Specifically, the power consumption of a single 5G macro BS is about 3500W, while the power consumption of a single 4G macro BS is about 1300W. Therefore, the power consumption of the former is about three times that of the latter. In addition, compared to 4G networks, 5G networks are operated at higher frequency bands, and the attenuation is more serious. To achieve the same coverage area as 4G networks, the number of 5G BSs is about three to four times that of 4G BSs, so the deployment costs are relatively high. Moreover, a large number of radio frequency (RF) chains and complex signal processing units are required adapting the massive multi-input multi-output (MIMO) architecture adopted in the 5G networks \cite{9113273}. Accordingly, compared with 4G BS, the design cost of 5G BS is also greatly increased. Therefore, it is imperative to seek a novel transceiver architecture that can reduce power consumption and cost for future beyond 5G and 6G networks.

Recently, reconfigurable meta-surface (RMS) has been proposed as a promising technology \cite{8811733}. In detail, RMS is a plane, which is composed of a large number of low-cost passive elements. The amplitude and phase shift of these elements can be adjusted independently by intelligent controller equipped with RMS, thereby generating directional beams. Due to the almost passive full-duplex operation mode, the RMS only passively reflects or transmits the incident signal. Therefore, compared with traditional relay technology, RMS does not encounter the problems of self-interference and additional noise. Compared with the existing multi-antenna technology, it does not need complicated RF chains and signal processing modules, so the required hardware cost and power consumption are much reduced. These advantages greatly show the potential of RMS in the next generation communication networks.

RMS is mainly divided into reflective mode and transmissive mode. At present, more work focuses on the two modes RMS-assisted communication \cite{9531372,9509394,li2021robust,9200683}. The RMS of the reflective mode or the transmissive mode realizes the reconstruction of the wireless environment by adjusting the amplitude and phase shift of the incident signal, realizing the enhancement of the desired signal strength and the reduction of the interference signal strength, thereby improving the spectral efficiency, energy efficiency and coverage area of the networks. In addition to being used for auxiliary communication, RMS can also be designed as a transmitter \cite{9133266,9570775}, which are two completely different research perspectives. The latter is also in its infancy, and it is very promising. It can greatly meet the requirements of future networks for low power consumption and low cost transceivers. However, compared to the reflective RMS multi-antenna system, the transmissive RMS can be designed to have a higher efficiency \cite{bai2020high,7448838}. This is mainly due to the following reasons. First, the transmissive RMS multi-antenna system has no feed blockage (occlusion). Second, the transmissive RMS multi-antenna system will not cause self-interference problems. The aperture efficiency of the transmissive RMS multi-antenna system is higher. Finally, the operating bandwidth of the transmissive RMS multi-antenna system is typically larger. The details will be explained in the next section. These advantages make the transmissive RMS multi-antenna system very promising for the future networks.

Although the transmissive RMS multi-antenna system has many advantages, it is composed of passive transmissive RMS and a single antenna feed, which is still very different from the traditional multi-antenna system composed of active components. So far, the transmissive RMS multi-antenna system is in its initial stage, it is crucial to depict its framework and mechanism. Therefore, this promotes this paper to give an overview of the transmissive RMS multi-antenna system, including its advantages, network architecture, transmission mechanism, modulation principle, channel model, channel estimation. Moreover, we present the design methodology of the downlink (DL) transceiver and the uplink (UL) transceiver. In addition, numerical results are provided to verify the effectiveness of the proposed algorithm in transmissive RMS multi-antenna systems. Finally, some potential research directions for the future work are given.
\begin{figure*} [t!]
	\centering
	\subfloat[\label{fig:a}]{
		\includegraphics[scale=0.20]{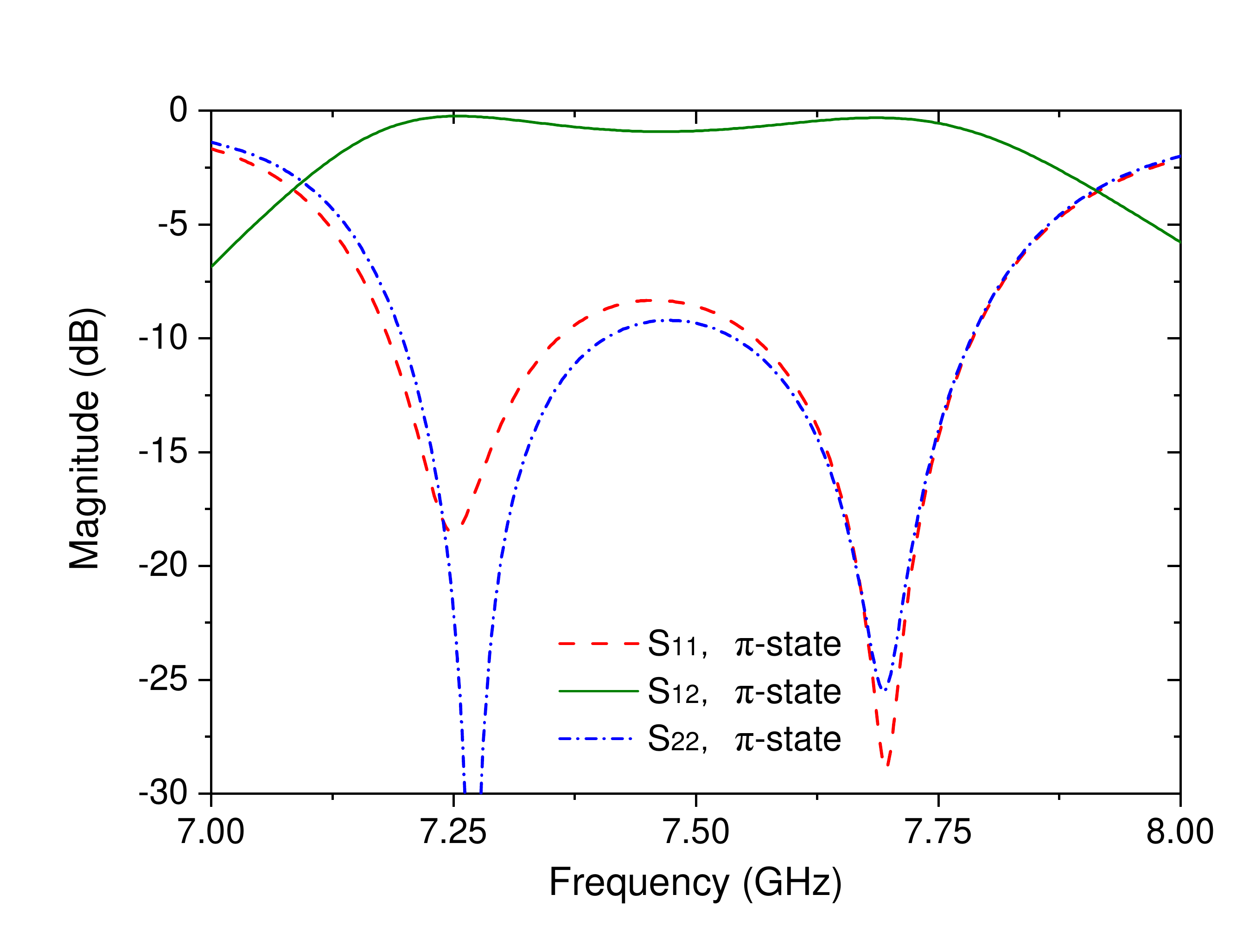}}
	\subfloat[\label{fig:c}]{
		\includegraphics[scale=0.20]{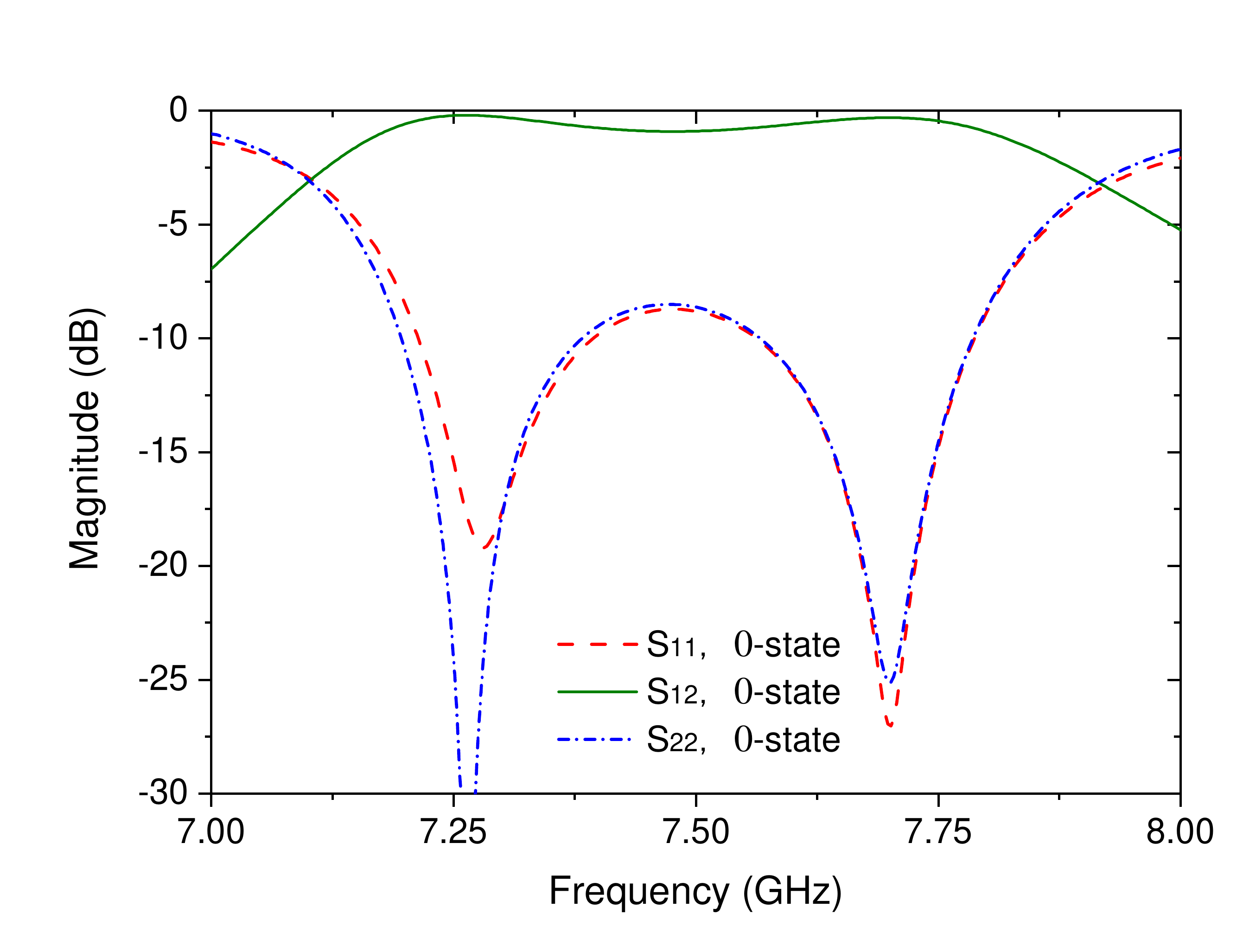}}
	\subfloat[\label{fig:e}]{
		\includegraphics[scale=0.20]{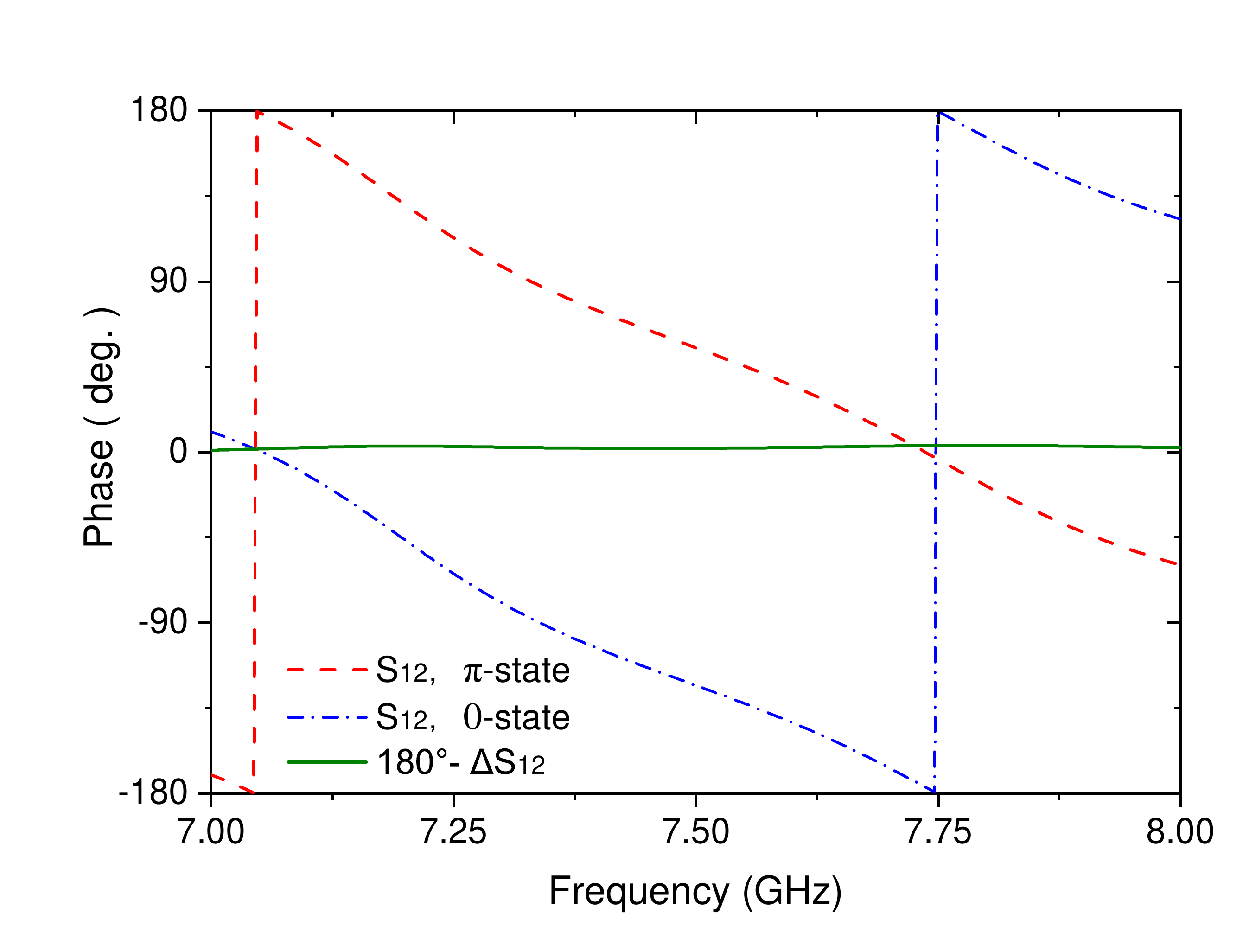}}
	\caption{Simulated scattering coefficients of the 1-bit transmissive unit. a) Magnitude for $\pi$-state. b) Magnitude for 0-state. c) Phases for both $\pi$/0 states \cite{bai2020high}.}
	\label{fig3} 
\end{figure*}
\section{The Advantages of Transmissive RMS Multi-Antenna System}
In this section, we first provide the advantages of our proposed transmissive RMS multi-antenna system compared to the traditional multi-antenna system and the reflective RMS multi-antenna system. 

Compared with the traditional multi-antenna system, since the transmissive RMS multi-antenna system is only composed of a single feed (receiving) antenna and passive transmissive RMS without complicated RF chains and signal processing units, its power consumption and cost can be greatly reduced. Then, we mainly elaborate on the advantages of the transmissive RMS multi-antenna system over the reflective RMS multi-antenna system from four aspects.
\subsection{No feed blockage (occlusion)} The user and feed antenna of the reflective RMS multi-antenna system are located on the same side of the RMS, while the user and receiving antenna of the transmissive RMS multi-antenna system are located on the opposite sides of the RMS. Therefore, there is no feed blockage (occlusion) problem in the transmissive RMS multi-antenna system, which will make the aperture efficiency of the transmissive RMS multi-antenna system higher. 
\subsection{No self-interference problem} For the reflective RMS multi-antenna system, the incident signals and the reflective signals are on the same side of the RMS, which will bring self-interference problems. As for the transmissive RMS multi-antenna system, the incident signals and the transmissive signals are on the opposite sides of the RMS, which can avoid this problem well. Therefore, the transmissive RMS multi-antenna system is more promising for the more complicated communication environments in the future.
\subsection{Higher aperture efficiency} According to the literature \cite{bai2020high} and \cite{7448838}, the aperture efficiency of the transmissive RMS multi-antenna system can be designed to be higher than that of the reflective RMS multi-antenna system. For instance, the literature \cite{bai2020high} gives the simulated scattering coefficient of the 1-bit transmissive unit as shown in the Fig. 1, where $S_{11}$ and $S_{12}$ represent reflective coefficient and transmissive coefficient, respectively. According to Fig. 1 (a) and (b), it can be seen that from 7.12 to 7.85 GHz, in both 0-state and $\pi$-state, the transmissive coefficient is above -2dB, and the reflective coefficient is below -10dB. Therefore, the transmissive RMS multi-antenna system can be designed to achieve higher aperture efficiency while avoiding the feed blockage (occlusion).
\subsection{Higher operating bandwidth} It can be seen from Fig. 1 (c) that the phase difference in transmissive coefficients of the $\pi$/0 states remains stable at around $\pi$ with very small deviation over a wide frequency band. However, the reflective phase difference of the reflective RMS is unstable enough \cite{liu2021multifunctional}. The more stable phase difference of the transmissive RMS multi-antenna system makes it have a larger operating bandwidth.

\section{Architecture and Transmission Mechanism of Transmissive RMS Multi-Antenna System}
As shown in Fig. 2, the architecture consists of a feed (receiving) antenna, RMS with $M$ transmissive elements, intelligent controller and $K$ users. In DL, the feed antenna can radiate single-frequency electromagnetic (EM) waves, while in UL, the receiving antenna can receive signals. Let ${f_{{m}}} = {\beta _{{m}}}{e^{j{\theta _{{m}}}}}$ denote the RMS transmissive coefficient, where ${\beta _{{m}}} \in \left[ {0,1} \right]$ and ${\theta _{{m}}} \in \left[ {0,2\pi } \right)$ respectively represent the transmissive amplitude and phase shift of the $m$-th element of RMS. Moreover, the controller equipped with RMS can adjust the amplitude and phase shift of the transmissive element to achieve the modulation and beamforming of the DL, as well as the enhancement of the UL incident signal. It is worth noting that due to the unique structure of the transmissive RMS multi-antenna system, DL can adopt multi-user space division multiple access (SDMA), and UL can adopt multi-user orthogonal frequency division multiple access (OFDMA) to achieve multi-user diversity. The specific DL and UL transmission mechanism is as follows:
\subsection{Downlink transmission} We have proposed the downlink transmission process of the transmissive RMS multi-antenna system in \cite{9570775}. Specifically, the information from the signal source is modulated by the controller (high-order modulation can now be achieved by time sequence modulation \cite{9133266}) as part of the RMS transmissive coefficient, and the beamforming design proposed by \cite{9570775} is used as the other part of the RMS transmissive coefficient. The two are multiplied as the RMS transmissive coefficient. In other words, the RMS controller can realize information modulation and beamforming design through the corresponding control signal, and the coefficient can be mapped to each element of the transmissive RMS. Then, the feed antenna sends a single tone signal to carry and send the information of each element to multiple users, and the multi-user diversity can be achieved by using SDMA with beamforming.
\subsection{Uplink transmission} This process has been studied in \cite{li2021uplink}, which is devoted to the performance improvement of the uplink transmission via RMS transmissive coefficient design for the transmissive RMS multi-antenna system. Since the receiver is equipped with a single antenna, we consider multi-user access by using OFDMA. The transmissive RMS strengthens the uplink signal and forwards it to the receiving antenna, and then the receiving antenna sends the received signal to the controller for demodulation and decoding.
\begin{figure}
	\centerline{\includegraphics[width=9cm]{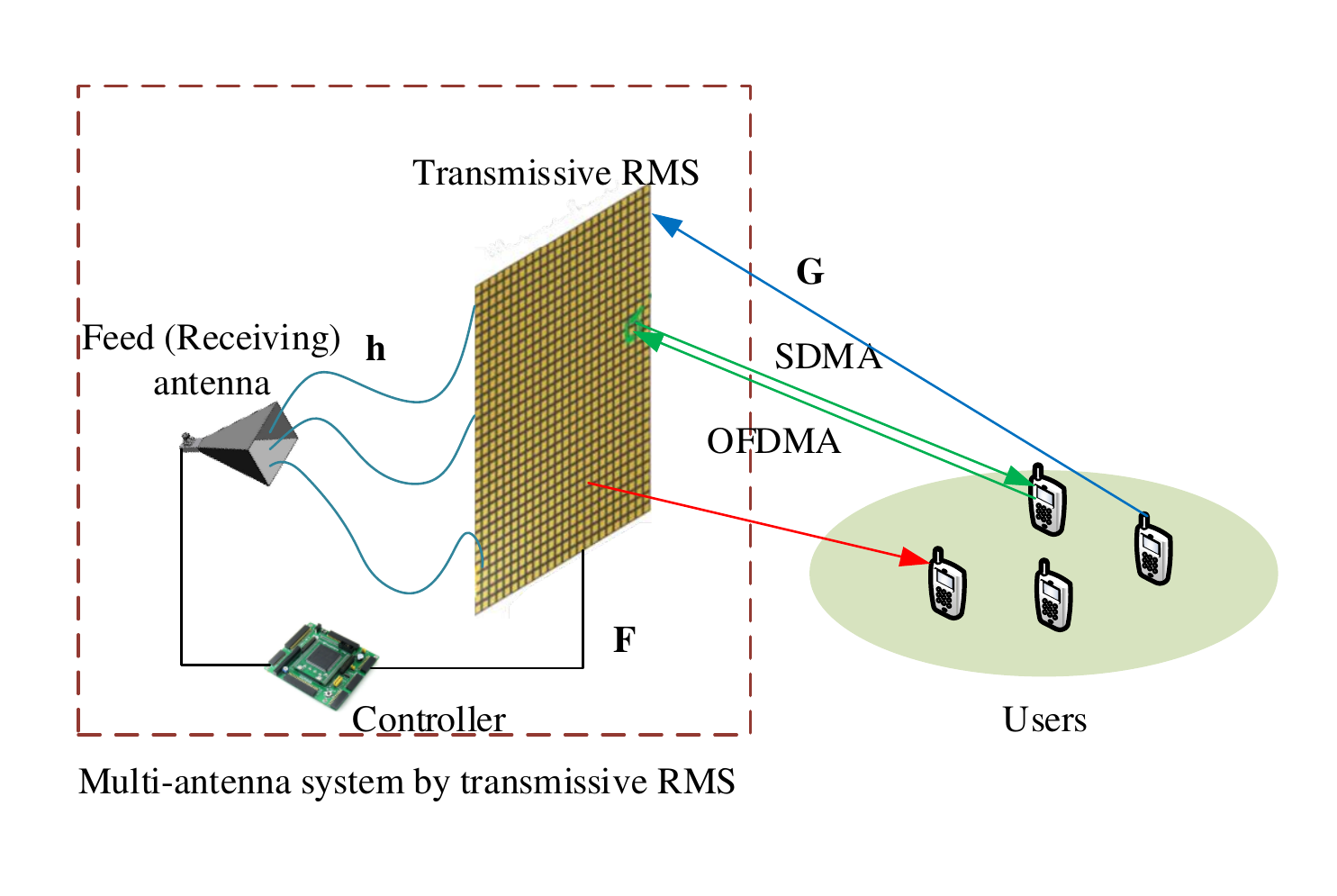}}
	\caption{Multi-antenna architecture by transmissive RMS.}
\end{figure}
\section{DL Modulation Principle}
In this section, we mainly provide the DL modulation principle of the transmissive RMS multi-antenna system. Note that the RMS in the transmissive RMS multi-antenna system does not have the complicated RF chains and baseband processing capabilities, so the conventional modulation method cannot be directly applied for this system. Several modulation schemes that can be used in the proposed system are described in detail below.

The modulation for this system is mainly divided into constant envelope modulation and non-constant envelope modulation. In the case of using constant envelope modulation, taking binary phase shift keying (BPSK) and quadrature phase shift keying (QPSK) as instances, we can control the conduction state of the diode in each element of the RMS by a control signal, i.e., BPSK and QPSK can be realized by 1-bit and 2-bit metasurfaces, respectively. Similarly, higher-order phase modulation can also be achieved in this way, but it is challenging. This is mainly due to space constraints, which is hard to accommodate a larger number of diodes in each RMS element. Moreover, the control signal needs to be multiplexed, which is also challenging to design. However, the degree of control freedom for constant envelope modulation remains at one. In this way, only one modulation symbol with a constant envelope can be generated on the carrier frequency, and high-order phase and amplitude joint modulation (i.e., non-constant envelope modulation) such as 16-quadrature amplitude modulation (QAM) cannot be realized, which will greatly limit the transmission rate. 

To solve this problem, a nonlinear modulation technique (i.e., time sequence modulation) can be used to achieve high-order phase and amplitude joint modulation \cite{9133266}. As shown as in Fig. 3, it is worth noting that after adopting time modulation, the waveforms of control signal has two degrees of control freedom, i.e., phase start time ${t_{on,n}}$ and conduction duration ${\tau _n}$. By expanding the Fourier series of the control symbol, we can get its amplitude and phase shift on the $l$-order harmonic.  The amplitude and phase shift of the $l$-order harmonics can be independently adjusted by setting the two degrees of freedom. Furthermore, the mapping relationship between the corresponding modulation scheme and the amplitude and phase of the $l$-order harmonic can be realized, i.e., the joint modulation of the phase and the amplitude can be realized.

Therefore, 16-QAM or higher order QAM can be achieved by applying the time modulation scheme, which can greatly improve the system performance. In addition, the above-mentioned BPSK, QPSK and higher order phase modulation can also be implemented. Moreover, for high-order modulation, which can be implemented on 1-bit or 2-bit RMS, the structure of the required RMS is simpler, and the control signal is also simpler. In addition, the scheme is reconfigurable, and the control signal waveform can be changed according to different modulation requirements. Therefore, the modulation scheme has great potential for the design of the proposed system.  
\begin{figure}
	\centerline{\includegraphics[width=9cm]{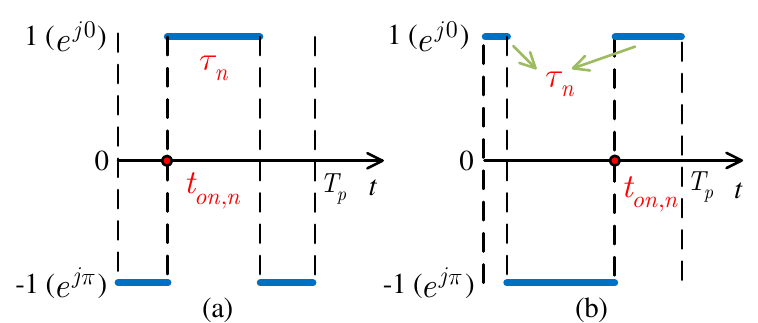}}
	\caption{The waveforms of control signal with two degrees of control freedom ${t_{on,n}}$ and ${\tau _n}$.}
\end{figure}
\section{Far-Near Field Channel Model}
As depicted in Fig. 4, the EM wave radiation field in the wireless communication network can be divided into far field and near field \cite{cui2021channel}, which have different channel models. The far-near field is determined by the Rayleigh distance $\frac{{2{D^2}}}{\lambda }$, where $D$ and $\lambda$ respectively denote array aperture and wavelength. The judgment method by applying Rayleigh distance is that, if the distance between the transmitter and the receiver is larger than Rayleigh distance, the channel can be regarded as a far-field channel, and the wavefront can be approximated as a planar wave. Otherwise, when the distance between the the transmitter and the receiver is less than Rayleigh distance, the channel can be modeled as near-field channel, and the wavefront is spherical wave. Hence, the channel model of the proposed RMS multi-antenna system can be elaborated as follows.
\begin{figure}
	\centerline{\includegraphics[width=9cm]{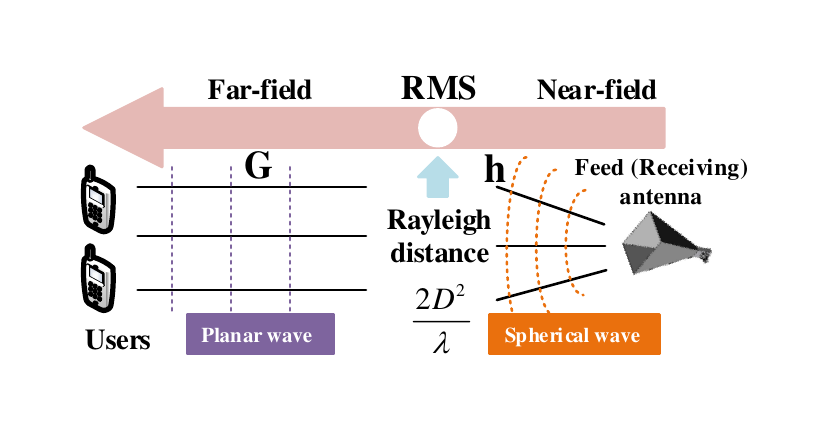}}
	\caption{Far-near field region separated by Rayleigh distance \cite{cui2021channel}.}
\end{figure}
\subsection{Far-field RMS-user channel model}
The channel from the user to the RMS can be named RMS-user channel. Since the distance from the user to the RMS is larger than the Rayleigh distance, the RMS-user channel can be regarded as a far-field channel, which is modeled under the assumption of planar waves. Considering that there are both line-of-sight (LoS) components and non-line-of-sight (NLoS) components between the user and the RMS, we model the RMS-user channel as a Rician fading channel. When the Rice factor is zero, the RMS-user channel can be transformed into a Rayleigh channel. Furthermore, each element of the NLoS component is an independent and identically distributed circularly symmetric complex Gaussian (CSCG) random vector with zero mean and unit variance. For the LoS component, it can be modeled in two ways: uniform linear array (ULA) and uniform planar array (UPA). It is obvious that the UPA is more practical due to the RMS is a plane.
\subsection{Near-field RMS-feed (receiving) channel model}
The channel from the RMS to the feed (receiving) antenna can be named RMS-feed (receiving) channel. Since the distance from the feed (receiving) antenna to the RMS is less than the Rayleigh distance, the RMS-feed (receiving) channel can be regarded as a near-field channel, which is modeled under the assumption of spherical waves. Considering that there is no blockage (occlusion) between the RMS and the feed (receiving) antenna, we can model it as a LoS channel with UPA. For the specific modeling and expression of the far-near field channel, please refer to \cite{li2021uplink}.

\section{Channel Estimation}
Since both the uplink and downlink transmissive coefficient designs require channel state information (CSI), the channel estimation for the proposed transmissive RMS multi-antenna system is vital. Hence, we interpret the channel estimation for the proposed system in this section. It is worth noting that the transmissive RMS can only transmit the signal passively, so the channel estimation of the system is not easy. To estimate the DL channel for RMS beamforming design, we consider the time-division duplexing (TDD) protocol, and estimate the UL channel. According to the channel reciprocity, the CSI of DL can be obtained. For UL transmission, because the receiving antenna has the receiving ability, and the RMS transmissive element has no receiving ability, we can only obtain the CSI of the UL cascaded channel by direct cascaded channel estimation, and cannot directly obtain the CSI of the separable channel. Nevertheless, only when we obtain the CSI of the UL RMS-user channel, the CSI of DL RMS-user channel can be obtained according to the channel reciprocity, and then the DL RMS transmissive coefficient can be designed. Moreover, the near-field RMS-receiving channel can be obtained by calculation or measurement, and it is usually fixed. Therefore, in the case of known cascaded CSI and near-field CSI, we can learn from some RMS separable cascaded channel estimation methods to obtain far-field RMS-user CSI. Below, we provide the two channel estimation methods in detail, i.e., direct cascaded channel estimation and separable cascaded channel estimation.
\subsection{Direct cascaded channel estimation}
For the direct cascaded channel, i.e., the multiplication of two parts of the channel, the existing work usually proposes some channel estimation protocol, and then uses different algorithms to estimate it \cite{8879620}. Direct cascaded channel estimation algorithms usually include message passing (MP) algorithm, channel correlation method, optimization-based channel estimation algorithm, decomposition, interpolation recovery, etc.. It is worth noting that the MP algorithm usually requires a sparse representation of the channel, and the method of considering the channel correlation can reduce the overhead of the pilot. We can consider the above-mentioned channel estimation method to obtain the UL cascaded CSI.

\subsection{Separable cascaded channel estimation}
For the separable cascaded channel estimation, some studies have proposed the semi-passive RMS channel estimation method and a fully passive RMS channel estimation method \cite{9361077}. For semi-passive RMS, it is equipped with a part of active elements to receive the transmitted signal from the transceiver or user, and then estimate the angle and path loss of the transceiver-RMS channel and RMS-user channel. However, this method needs to feed back to the transceiver and the user for beamforming design again, which will reduce efficiency. In addition, it will also increase the cost of hardware. Full-passive RMS can solve this limitation well. For the full-passive RMS, the user transmits different pilot sequences to the transceiver through the RMS, and through the analysis of the received signal, a two-stage method can be used to estimate the cascade channel. Note that the near-field channel of the transmissive RMS multi-antenna system can be calculated or measured, and the cascaded channel can be obtained by the method in the previous subsection. Therefore, we can use the method mentioned above to estimate the RMS-user channel.

\section{DL Transceiver Design and Optimization}
With the UL transmission mechanism and CSI based on the above channel estimation, the DL transceiver can be designed and optimized. In \cite{9570775}, we have proposed a transmissive RMS transmitter architecture, and conducted a preliminary study on the downlink multi-user beamforming design (i.e., RMS transmissive coefficient design) and power allocation under the transmissive RMS transmitter architecture. Specifically, a transmitter design based on a transmissive RMS equipped with a feed antenna is first proposed, where the RMS array is distributed in UPA and we use a geometric channel model with $L$ propagation paths to model the channel between the transmitter and the user. Also, the modulation method is introduced in detail. Then, the achievable system sum-rate are maximized by jointly optimizing the RMS beamforming design and power allocation. Considering that the optimization variables are coupled, this formulated optimization problem is non-convex. Accordingly, we propose an alternating optimization (AO) technique based on difference-of-convex (DC) programming and successive convex approximation (SCA) to solve this problem. Via applying this algorithm, the non-convex optimization problem is transformed into a convex optimization problem, which can be effectively solved by using the CVX toolbox.

For DL transceiver design and optimization, we consider a system setup of $K=4$ users and $M=25$ RMS elements. We assume that the parameters of all users are the same. We compare the performance of the proposed algorithm with other baseline algorithms in Fig. 5. 
\begin{itemize}
\item Equal allocation (EA): The beamforming design is the same as our proposed algorithm, and the power allocation is equal allocation. 
\item Zero-forcing (ZF) beamforming: The beamforming design applies ZF algorithm, and the power allocation is equal allocation. \item Random allocation (RA): The beamforming design and power allocation adopt random allocation. 
\end{itemize}
Fig. 5 depicts the change of system sum-rate with the number of RMS elements. It can be seen that as the number of RMS elements increases, the system sum-rate increases. This is because more RMS elements can bring more spatial diversity gains. Meanwhile, more RMS elements can get more signal scattering paths, so that the user can get a stronger signal. In addition, when the number of RMS elements is the same, our proposed algorithm has a better gain than other baseline algorithms.
\begin{figure}
	\centerline{\includegraphics[width=8cm]{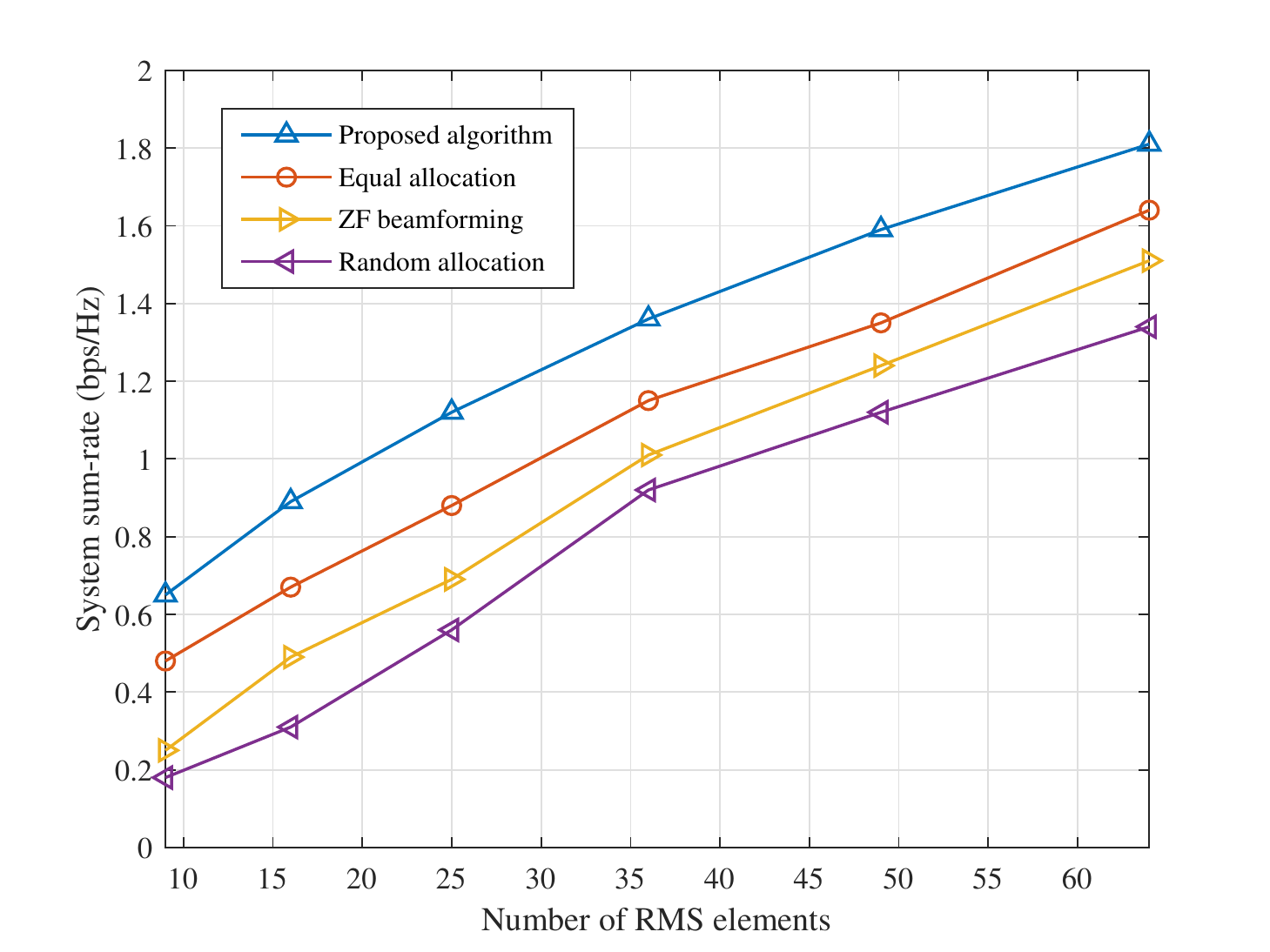}}
	\caption{DL system sum-rate versus number of RMS transmissive elements under different algorithm.}
\end{figure}
\section{UL Transceiver Design and Optimization}
After obtaining the uplink cascaded CSI through the above channel estimation algorithm, we have already studied uplink transceiver design and optimization for transmissive RMS multi-antenna systems in \cite{li2021uplink}. In detail, we propose a transmissive RMS-based receiver equipped with a single receiving antenna, and a far-near field channel model based on planar waves and spherical waves is given. Also, we discuss the advantages of the proposed architecture. In this architecture, the multi users access the networks by OFDMA due to the receiver is equipped with single antenna. Then, in order to maximize the system sum-rate of uplink communications, we formulate a joint optimization problem over subcarrier allocation, power allocation and RMS transmissive coefficient design. Due to the coupling of optimization variables, the optimization problem is non-convex. We apply the AO algorithm baesd on Lagrangian dual decomposition method, DC programming, SCA and penalty function methods to tackle this problem. Moreover, the superiority of the proposed scheme is verified by numerical simulation.

For UL transceiver design and optimization, we consider a 3D coordinate system setup with $K = 5$ users and $M=25$ transmissive elements. In order to evaluate the performance of the proposed algorithm, we consider the following benchmark algorithm. 
\begin{itemize}
	\item Benchmark 1 (three-stage algorithm): The algorithm does not perform alternate iterations after optimizing subcarrier allocation, power allocation and RMS transmission coefficient once. 
	\item Benchmark 2 (random coefficient algorithm): The algorithm optimizes subcarrier allocation and power allocation by Lagrangian dual decomposition method, and uses a random scheme for the RMS transmissive coefficient. 
	\item Benchmark 3 (random allocation algorithm): The algorithm uses a random scheme for subcarrier allocation, power allocation, and RMS transmissive coefficient design. 
\end{itemize}
 Fig. 6 shows the change of system sum-rate under different algorithms when the number of RMS transmissive elements increases. We can see that the performance of the proposed algorithm outperforms other benchmark algorithms. Moreover, the system sum-rate increases as the number of RMS transmissive elements increases. This is because the RMS transmissive element will strengthen the user’s uplink transmission signal and bring a certain gain, which is positively correlated with the number of RMS transmissive elements. Therefore, the more RMS transmissive elements, the more gain it brings, and the greater the system sum-rate. It provides a good guide for multi-antenna systems based on transmissive RMS, which can improve system performance by increasing the number of low-cost transmissive elements, which has great potential for the future networks that needs to reduce cost and power consumption.
\begin{figure}
	\centerline{\includegraphics[width=8cm]{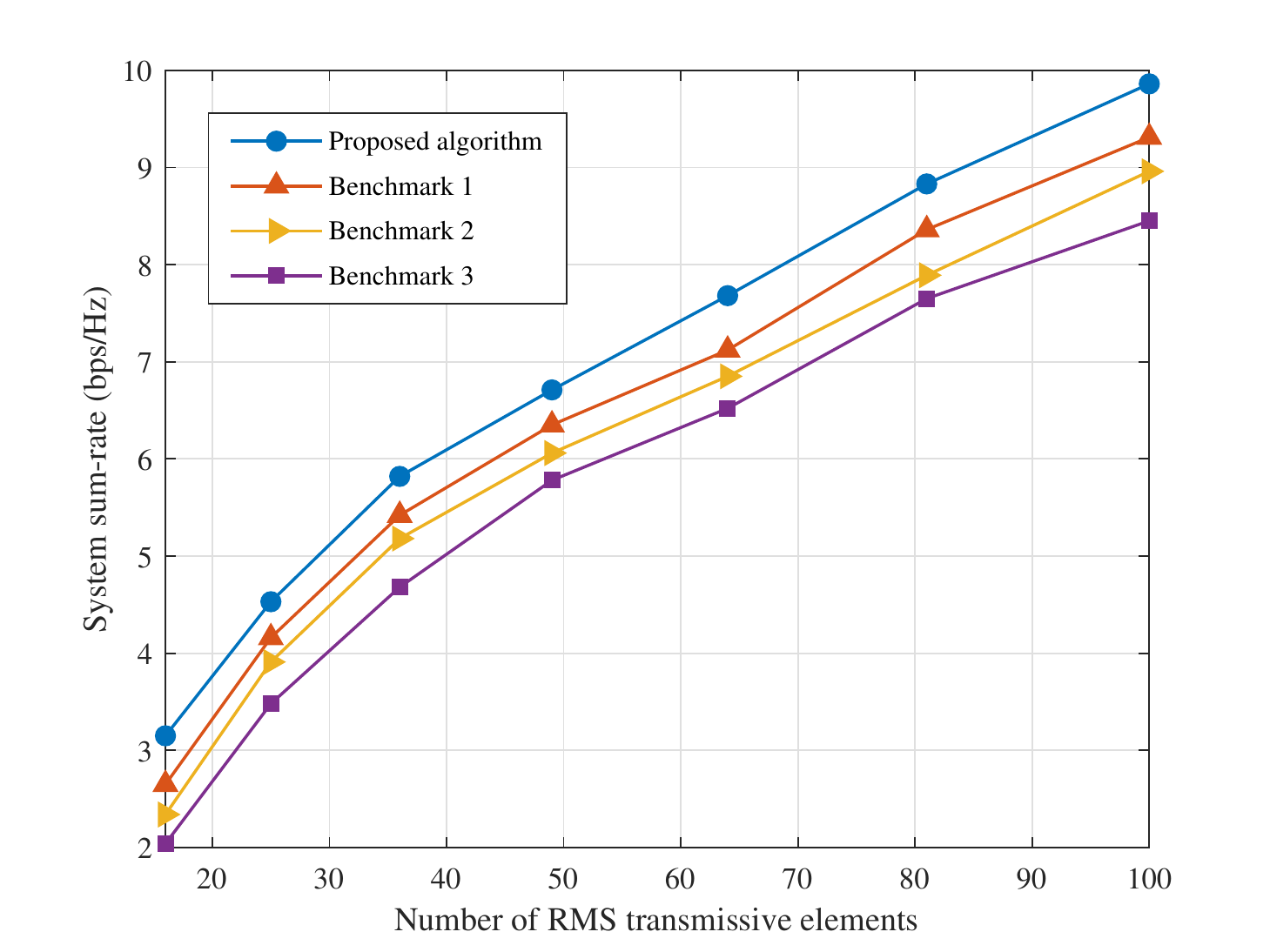}}
	\caption{UL system sum-rate versus number of RMS transmissive elements under different algorithm.}
\end{figure}

\section{Potential Directions}
With the advantages of the transmissive RMS multi-antenna system, in this section, we provide a potential research direction in the future under this new type of multi-antenna system.
\subsection{Non-orthogonal multiple access (NOMA) with transmissive RMS multi-antenna system}
For the UL of the transmissive RMS multi-antenna system, the receiver is a single antenna. Therefore, in addition to considering the orthogonal multiple access (OMA) scheme, NOMA is also a very important access scheme. Moreover, for DL, NOMA can well meet the large-scale connections of 6G networks and the higher service requirements of users. Unlike traditional OMA, NOMA can support multiple users to share the same resources, such as time, frequency, coding, etc.. Specifically, taking NOMA in the power domain as an example, the transmitter can use the same resources to serve multiple users, which can greatly improve the spectrum efficiency to meet the user's service requirements. For NOMA transmission in the DL and UL power domain, superposition coding and serial interference cancellation (SIC) technologies are applied at the transceiver and the user, respectively. By applying SIC technology, users with stronger channel gains can remove co-channel interference caused by users with weaker channel gains before decoding. It is a very novel work to consider the design of SIC decoding order and RMS transmissive coefficient at the same time.
\subsection{Robust beamforming design with transmissive RMS multi-antenna system}
At present, the design and optimization of DL and UL transceivers have been completed under the assumption that perfect CSI can be obtained. However, it is more practical to consider the design and optimization of DL and UL transceivers in the case of imperfect CSI. Specifically, we can consider the optimization of system performance when the channel estimation has a certain error. The modeling of channel estimation errors can be modeling based on bounded and based on statistical outage probability. When we consider that the channel estimation has a certain error, the solution of the optimization problem will become more complicated and challenging, but it is more practical, which can be used as a very meaningful work in the future.
\subsection{Wireless power transmission (WPT) with transmissive RMS multi-antenna system}
Considering that there will be more large-scale, low-power, energy-constrained Internet-of-things (IoT) devices in 6G networks, to provide continuous information transmission and energy transmission requirements, simultaneous wireless information and power transfer (SWIPT) and wireless powered communication networks (WPCN) can serve them well. We can propose WPT with transmissive RMS multi-antenna system to realize a new network paradigm for 6G WPT with lower power consumption and cost. Meanwhile, considering the limitation of system performance degradation due to distance limitation, we can also consider using reflective RMS to further strengthen, greatly improving the transmission efficiency and coverage, so as to realize the transmissive-reflective RMS-empowered WPT network architecture in 6G networks. However, the joint design for WPT network parameters and RMS coefficients is still challenging.
\section{Conclusions}
In this paper, we provide an overview of the efficient transmissive RMS multi-antenna system for achieving low power consumption and low cost transceiver in the beyond 5G and 6G networks. RMS can adjust its transmissive coefficient to achieve DL beamforming and UL signal enhancement. Since the design and optimization of transmissive RMS transceivers are new and have not been explored to a large extent, it is hoped that this paper can provide some useful guidance for future research. Meanwhile, we can foresee that the application of low-cost and low-power RMS in future wireless communication networks will cover the overall closed-loop wireless transmission link composed of transmitters, auxiliary wireless communications, and receivers. The new RMS-based wireless communication system is expected to achieve the best overall performance through joint optimization design.

\bibliographystyle{IEEEtran}
\bibliography{reference}
\begin{IEEEbiographynophoto}{Zhendong Li}
is currently pursuing his Ph.D. degree at Broadband Access Network Laboratory, Department of Electronic Engineering, Shanghai Jiao Tong University, China. His research interests include reconfigurable meta-surface (RMS) and unmanned aerial vehicle (UAV) communications.
\end{IEEEbiographynophoto}
\begin{IEEEbiographynophoto}{Wen Chen}
	is a professor at Broadband Access Network Laboratory, Department of Electronic Engineering, Shanghai Jiao Tong University, China. His interests include reconfigurable meta-surface, wireless AI, multiple access, coded cooperation, and green heterogeneous networks.
\end{IEEEbiographynophoto}
\begin{IEEEbiographynophoto}{Chong He}
	is an assistant professor of the Department of Electronic Engineering, Shanghai Jiao Tong University, China. His research interests include phased arrays, reconfigurable intelligent surface, DOA estimation, wireless location and multiple access wireless communications.
\end{IEEEbiographynophoto}
\begin{IEEEbiographynophoto}{Xudong Bai}
	is an associate professor of the School of Microelectronics, Northwestern Polytechnical University, China. His current research interests include phased arrays, electromagnetic metamaterials and metasurfaces, and OAM-EM wave propagation and antenna design.
\end{IEEEbiographynophoto}
\begin{IEEEbiographynophoto}{Jianmin Lu}
	 is currently the Executive Director of Huawei Wireless Technology Laboratory. His current research interest is in the area of signal processing, protocol, and networking for the next generation wireless communication.
\end{IEEEbiographynophoto}
%
%
%
%
%
%
%
%

\end{document}